\def\astroph{1}

\ifnum\astroph=0
\documentclass[12pt,preprint]{aastex}
\usepackage{aastexug}

\else
\documentclass{aastex}
\usepackage{emulateapj5}
\usepackage{apjfonts}
\usepackage{calc}
\fi

\shorttitle{Bright loops in the solar corona}
\shortauthors{B.V. Gudiksen \& {\AA}. Nordlund}

\newcommand{\Fig}[1]{Fig.\ \ref{#1.fig}}

\newcommand{\Sec}[1]{Section \ref{#1}}

\newcommand{\cf}{{cf.}}

\newcommand{\JJ}{{\bf J}}
\newcommand{\BB}{{\bf B}}

\ifnum\astroph=0
\newcommand{\FIGI}{
 \begin{figure}[t]
 \figurenum{1}
\plotone{f1.eps}
\caption{%
   Histogram of current density squared as a function
   of height. Dark colors are higher filling factor. The upper and
   lower line are the horizontal average of the magnetic field
   strength squared and the Joule heating squared, respectively.
 }
 \label{j2distr.fig}
 \end{figure}
}

\newcommand{\FIGII}{
\notetoeditor{It is here crucial that fig2a and b are stacked
  immediately above one another. Only one caption is needed,
  and only one label has been assigned (to 2b).}
 \begin{figure}[t]
 \figurenum{2a}
\plotone{f2a.eps}
\end{figure}
\begin{figure}[t]
\figurenum{2}
\plotone{f2b.eps}
 \caption{
 Synthetic TRACE 171 emission measure,
 averaged over the z-direction, and raised to the power
 0.5, to soften the contrast(top).
 SOHO/MDI magnetogram of AR 9114, used as initial
 condition, and the loops from \Sec{sec:results} (bottom).
 One and a half box width is shown.
 }
 \label{trace171.fig}
 \end{figure}
}

\newcommand{\FIGIII}{
 \begin{figure}[t]
 \figurenum{3}
\plotone{f3.eps}
 \caption{
  Temperature (full line), gas pressure (dashed line), and
  current helicity ($\alpha=\JJ\cdot\BB/B^2$ -- dotted line) as a function of height $x$
  for the two loops traced
  in \Fig{trace171}.
  }
 \label{brightloop.fig}
 \end{figure}
}
\else

\newcommand{\FIGIem}{
 \figurenum{1}
 \centerline{\includegraphics[width=8.2cm]{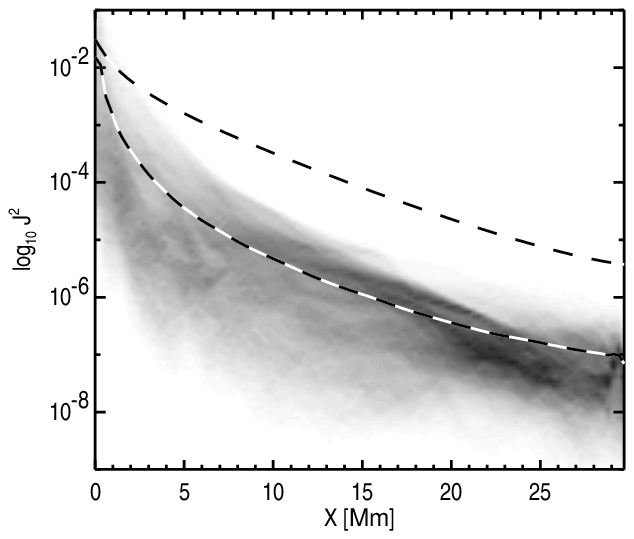}\hspace*{0.5cm}}
 {\small \noindent Fig. 1:
   Histogram of current density squared as a function
   of height. Dark colors are higher filling factor. The upper and
   lower dashed lines are the horizontal averages of the magnetic field
   strength squared and the Joule heating squared, respectively.
 }
 \vspace{10pt}
 \label{j2distr.fig}
}

\newcommand{\FIGIIem}
{
 \figurenum{2}
 \vspace*{0.3cm}
 \centerline{\includegraphics[width=8.2cm]{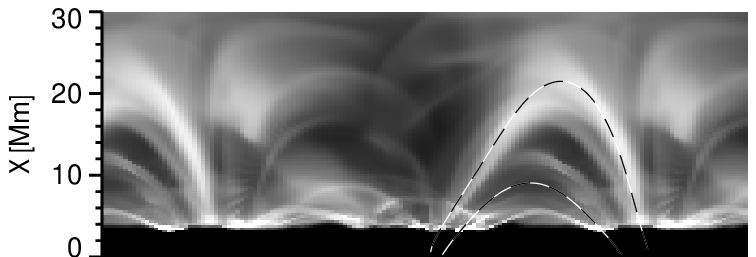}}
 \centerline{\includegraphics[width=8.2cm]{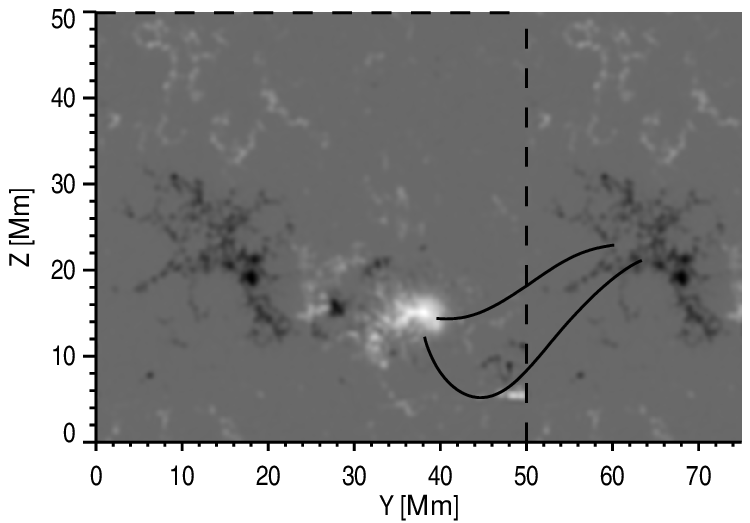}}
 \vspace*{0.2cm}
 {\small \noindent Fig. 2:
 Synthetic TRACE 171 emission measure,
 averaged over the z-direction, and raised to the power
 0.5, to soften the contrast(top).
 SOHO/MDI magnetogram of AR 9114, used as initial
 condition, and the loops from \Sec{sec:results} (bottom).
 One and a half box width is shown.
 }

 \vspace*{10pt}
 \label{trace171.fig}
}

\newcommand{\FIGIIIem}{
 \figurenum{3}
 \centerline{\includegraphics[width=8.2cm]{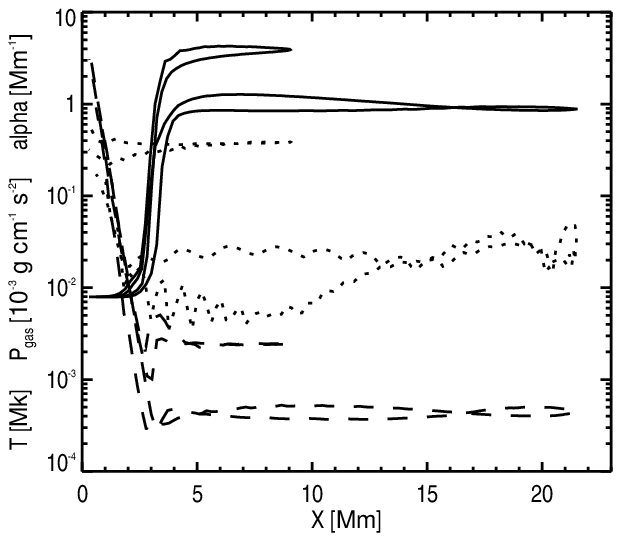}}
 {\small \noindent Fig. 3:
  Temperature (full line), gas pressure (dashed line), and
  current helicity ($\alpha=\JJ\cdot\BB/B^2$ -- dotted line) as a function of height $x$
  for the two loops traced
  in \Fig{trace171}.
  }
 \vspace{10pt}
 \label{brightloop.fig}
}
\fi

\begin{document}

\title{Bulk heating and slender magnetic loops in the solar corona}

\ifnum\astroph=1
\author{B.V. Gudiksen and {\AA}. Nordlund}
\else
\author{Boris Vilhelm Gudiksen}
\affil{The Institute for Solar Physics}
\affil{SCFAB\\ 10691 Stockholm\\ Sweden}
\email{boris@astro.su.se}
\and
\author{{\AA}ke Nordlund}
\affil{Astronomical Observatory}
\affil{NBIAfG, Copenhagen University}
\affil{{\O}ster Voldgade 3\\1350 Copenhagen K\\Denmark}
\email{aake@astro.ku.dk}
\fi


\begin{abstract}
The heating of the solar corona and the puzzle of the slender high
reaching magnetic loops seen in observations from the Transition
Region And Coronal Explorer(TRACE) has been investigated through 3D
numerical simulations, and found to be caused by the well
observed plasma flows in the photosphere displacing the foot points of
magnetic loops in a nearly potential configuration. It is found that
even the small convective displacements cause magnetic dissipation
sufficient to heat the corona to temperatures of
the order of a million Kelvin. The heating is intermittent in both space
and time---at any one height and time it spans several orders of magnitude,
and localized heating causes transonic flows along field lines, which
explains the observed non-hydrostatic stratification of loops that are
bright in emission measure.
\end{abstract}
\section{Introduction}
The Sun has provided us with a problem that has puzzled researchers
for many decades. The solar corona has a sustained temperature
of the order of one million Kelvin, three orders of magnitude higher than
the photosphere. On top of that the corona shows magnetic structures
in the form of loops reaching high into the corona through several
pressure scale heights while still being below the resolution limit of
the best instruments available at the appropriate wavelengths. Even
though several heating processes could be at work, with bulk heating
largely independent of the mechanism creating the slender loops, both effects
are concentrated around active regions, and there are therefore strong
indications that they are related.  If the bulk and loop heating
mechanisms are actually the same,
the heating mechanism must operate partly on scales smaller than the
resolution limit of 1.0\arcsec ($\sim$ 725 km) of the TRACE instrument
\citep{Aschwandenetal00} in order to explain the high reaching slender
loops, and partly on scales comparable to the size of an active region.
Both the bulk
heating and the loops are believed to be related to magnetic
processes, but identifying the main contributing effects has
proven difficult.

Coronal active regions need a continuous thermal
energy input of $10^{6}-10^7\, {\rm{ergs \, cm^{-2} \, s^{-1}}}$
in order to counter thermal conduction and X-ray losses at
coronal temperatures. Several heating
mechanisms have been proposed, among them wave dissipation, direct
current (DC) dissipation, and nano--flares. Wave dissipation is only
possible for Alfv{\'e}n waves, since the other magneto--sonic wave modes are
diffracted and dissipated by the strong wave speed gradient in the
chromosphere region. To make Alfv{\'e}n waves dissipate their energy in
the corona is not easy 
and the physical requirements are hard to meet.
Direct current dissipation is relatively
easy to realize, but it has been unknown whether it is possible to
induce enough current in the observed, nearly potential magnetic field
configuration, and at the correct heights. Nano--flares have until
recently been a promising
candidate, but observations now seem to indicate that the power in the
observed nano--flares is insufficient \citep[][and references
  therein]{Aschwandenetal00,Parnelljubb00}.

The DC heating mechanism appears to be the most promising one.  It was
proposed 30 years ago \citep{Parker72}, and has received a lot of
attention over the years \citep[e.g.][to mention just a few]{Parker72,
  Parker83, Sturrock+Uchida81, vanBall86, Mikic+ea89,
  Heyvaerts+Priest92, Longcope+Sudan94, Galsgaard+Nordlund95xc,
  Hendrix+96, Gomez+00}.
These works have
generally argued that the mechanism is feasible, but have been unable
to actually demonstrate
that it works, in the sense that it produces the right amount of
heating, and the observed type of coronal structures.  An important
step forward was taken when \citet{Galsgaard+Nordlund95xc} and
\citet{Hendrix+96} showed that the dissipation does not depend on,
or depends only very weakly on the magnetic Reynolds number (or
equivalently, the resolution of numerical experiments).

As pointed out by \citet{Aschwanden01c} and others, it is necessary
to deal with the correct geometry and stratification in order to
verify or falsify a proposed mechanism.  One cannot, for example,
hope to reproduce the distribution of heating along the length of
loops without representing the loop geometry reasonably correctly.
In addition, since the main cooling mechanisms depend strongly on
topology (for thermal conduction) and on density stratification
(for radiative cooling), it is obvious that the question of coronal
heating can only be answered by employing a sufficiently realistic setup.

We use a 3D magneto-hydro-dynamics (MHD) code
\citep{Nordlund+Galsgaard95mhd}
to simulate a typical
scaled down active region, including a simple photospheric driving and
starting with a potential field extrapolation from an MDI
magnetogram, in order to investigate if the convective motions of
the solar photosphere are sufficient to heat the corona through
magnetic field line braiding. We conclude that even though the
magnetic field remains not far from a potential field
configuration, the convective driving is all that is needed to
heat the corona and produce hot loops such as those seen in,
e.g., the TRACE 171 {\AA} filter, through a DC heating mechanism.

\section{Numerical Procedures}\label{code}

The numerical code uses 6'th order differential operators and 5'th
order translational operators on a staggered mesh to solve the fully
compressible MHD equations.
Radiative cooling \citep{Kahn76} and Spitzer conductivity
\citep{spitzer56} along the magnetic field are included in the energy equation.

The velocity field at the lower boundary is updated  through a procedure
that smoothly changes horizontal velocities for a given scale from
one random pattern to another over a turn-over time appropriate
for each scale.
The horizontal velocity pattern is generated from a
velocity potential with randomly phased 2-D Fourier components,
with amplitudes that follow a power law $k^{-p}$.  The velocity
power spectrum is then $P(k) \propto k^{3-2p}$, typical velocities
at scales $1/k$ are $v(k) \propto \sqrt{k P(k)} \propto k^{2-p}$,
and the corresponding turn-over times $\tau(k) = 1 / k v(k)
\propto k^{p-3}$.

We choose to set $p=1$, which is consistent with observed super granulation
and granulation turnover times; $\sim 30$ hours at scales $\sim 30$ Mm, and
$\sim 1000$ s at scales $\sim 3$ Mm, respectively.  A power spectrum
$P(k) \propto k$ is also consistent with the large scale part of the
power spectrum in well established simulations of convection on granular and
meso granular scales \citep{stein98}.  At
smaller scales the granulation velocity power deviates from $P(k) \propto k$,
peaking at $k \sim 4-6 \,{\rm Mm}^{-1}$ and decreasing at even larger $k$.
These scales are, however, below the horizontal resolution of the
present experiment.

We increased the velocity field amplitude by about a factor two,
relative to the rms horizontal velocity from convection
simulations, in order to counter the effects of our extended
chromosphere and transition region (see \Sec{initialcond}).
It remains to be seen if
future simulations, with improved vertical and horizontal
resolution, will be able to achieve similar results while using a
properly normalized velocity power spectrum.

\section{Initial conditions}\label{initialcond}
The initial conditions were resolved on a uniform grid with
111 points in the vertical direction $(x)$, including 11 ``ghost zones'',
and $100 \times 100$ grid points in the horizontal (periodic)
directions $(y,z)$. The grid spans a volume $30\times 50\times 50\,
\rm{Mm}^3$ excluding ghost zones, giving a resolution of 0.3 Mm
vertically and 0.5 Mm horizontally.

The requirements of having a high plasma beta
($\beta=P_{gas}/P_{mag}$) at the lower boundary, while at the
same time reaching the much lower coronal pressures across a
thin chromosphere, gives a resolution problem not
easily solved on a uniform grid, and requires a compromise.
We chose to extend the thickness of the ``chromosphere'' to about
4 Mm, in order to be able to cover the large change of pressure
there with about one point per scale height.

This solution has the side effect that if the heating
decreases exponentially with height \citep[as seems to be the
case; see ][]{Schrijver99,Aschwanden01a,Aschwanden01b}, we
underestimate the heating.
Our chromosphere is roughly 1.5 Mm thicker than the VAL models
of the solar chromosphere \citep{VAL81}.  The corresponding decrease
of the heating is approximately compensated
for by our increase in velocity driving amplitude.

The lower
boundary and photosphere is kept at a constant temperature
$8\times 10^3 \,{\rm{K}}$.
The upper boundary is kept at the initial temperature ($10^6 \,{\rm{K}}$)
during the start-up phase. Thereafter a vanishing vertical
derivative of temperature is assumed, thus enforcing a vanishing
vertical component of the thermal conductive flux there.

The initial condition for the
magnetic field is derived from a magnetogram
of active region 9114, observed near disc
center by TRACE and by the Michelson Doppler Imager (MDI) on the SOlar and
Heliospheric Observatory (SOHO) on August 8, 2000.
The original magnetogram was cropped at a perimeter chosen
to intersect a minimum of magnetic field, and was then made
periodic by taking a Fourier transform. 
The physical range of the observed
active region was then scaled down from $\sim 250$ Mm to 50 Mm in
order to fit in the computational domain, while still having magnetic
structures on even the smallest scales. This was done because the
details of the magnetic
field are not terribly important, since we did not try to
reenact the dynamics of AR9114 in particular.  We are satisfied
to have a reasonable realistic initial distribution of the magnetic
field in our model, and aim rather at studying the generic behavior
of active regions.

The initial magnetic field was obtained by making a potential extrapolation
from the vertical magnetic field from the magnetogram.  In the
subsequent evolution, the magnetic field at the boundary evolves
under the control of the random horizontal velocity field,
specified as explained above.  The vertical velocity is assumed
to vanish at the boundary.  The horizontal velocity field is
divergence free by construction, so the evolution corresponds to
moving the foot points of the magnetic field around, while
conserving the magnetic flux density.

\section{Results}\label{sec:results}

After an initial start-up phase the simulation evolves
towards a quasi-stationary configuration with a hot tenuous corona
with temperatures of the order a million K.  The horizontally
averaged temperature peaks at about $1.3$ MK, about 6 Mm above
the transition zone. Maximum temperatures $\sim 4$ MK are reached
about 10 Mm above the transition zone, where the average temperature
is $\sim 1.1$ MK (these are mentioned merely as examples---the
detailed numbers are expected to vary between active regions).

The temperature, density, pressure, electric current density,
and velocity all vary considerably across horizontal planes,
but tend to vary much less dramatically along magnetic field
lines.

The average density at a height of 10 Mm above the transition
zone is $\sim 2\times10^{-15}$ g\,cm$^{-3}$, or about $10^9$ atoms per
cm$^{-3}$.  The maximum density is typically 20--30 times higher; i.e.,
around 2--3$\times 10^{10}$ cm$^{-3}$.  The large variation of density
and temperature between magnetic field lines is the main
cause of the appearance of loop--like structures in \Fig{trace171}.

The chromosphere and transition region turn out to be of crucial
importance, as predicted by \citet{Aschwanden01c}.
The root mean square electric current decreases exponentially by
almost three orders of magnitude in the lower chromosphere.  There,
the magnetic field is non--force--free, with a very intermittent
Lorentz force that interacts with both gas pressure gradients and
inertial forces.

In the upper chromosphere the magnetic field gradually becomes
nearly force--free, and therefore the scale height of the
$J^2$ distribution approaches that of $B^2$ (\cf~\Fig{j2distr}).
Even though there are large variations in horizontal planes,
over which the current distribution is very intermittent, the
height dependence of the root mean square electric current in
the corona more or less follows that of the magnetic field, as
is to be expected from a distribution with approximately
``constant alpha'' ($\alpha = \JJ\cdot\BB/B^2$) along magnetic
field lines.

In the transition region between the chromosphere and the corona
the density and pressure scale heights rapidly change to much
larger values.  The average Joule dissipation thus decreases
faster than the radiative cooling in the corona, and therefore
the peak of the average temperature occurs low in the corona.
The temperature structure from the transition region to the maximum
of the temperature is mainly determined by heat conduction along
magnetic field lines.

The average Joule dissipation, which balances the sum of
thermal conduction losses and radiation losses, increases
monotonically with decreasing height, which makes it difficult
to unambiguously define an average coronal heating rate.
However, over the region in height where the temperature is
larger that one million K, the average heating rate is
$\sim 2\,10^6$ erg cm$^{-2}$ s$^{-1}$.

The heating varies considerably over the horizontal
plane, with much larger average rates in the immediate
neighborhood of the active region patches of strong magnetic
field (except right above the sunspot, where the heating rate
is low).  The heating rate in the 25\% of the horizontal
area that covers the central part of the active region
is 2--3 times higher than the average over the whole
model.

\ifnum\astroph=1
\FIGIem
\fi

The Joule heating is plotted as a scatter histogram in \Fig{j2distr}.
Although there is a relatively large scatter in the numerical values
at each height, the horizontal average shows a smooth,
roughly exponential height dependence. The magnetic dissipation thus
indeed decreases roughly exponentially with height,
as was proposed by \citet{Schrijver99}, and as was deduced
from TRACE data by \citet{Aschwanden01a}.

Note that the scale height of the heating empirically is found to
increase with loop length; \citet{Aschwandenetal00} report a
ratio $\sim 0.2$ between the scale height and the loop length.
Such an approximate proportionality
should indeed be expected, if the heating is controlled by a
roughly constant winding number from one loop
end to another \citep{Galsgaard+Nordlund95xc}.
\ifnum\astroph=1
\FIGIIem
\noindent
\fi

Slender loops of the type observed by TRACE in full size active regions
are seen in this simulation as
well (\Fig{trace171}). They show that it is possible, merely
by using a random photospheric convective velocity pattern,
to create thin slender loops that do not expand much with height,
and are almost isothermal, as required
if they are to be observed in the narrow TRACE filters. Two loops have
been followed; the first was selected for its brightness in the TRACE
171 filter and 
the second for its large electric current density. Their magnetic field line
traces may be seen in \Fig{trace171}, and their
temperature, gas pressure and current helicity $\alpha$ have been plotted
along the loops in \Fig{brightloop}. The loops both show a nearly
constant pressure and temperature in the corona, maintained by the
Spitzer conductivity forcing a small gradient in temperature along the
loops.  The winding number for the small loop is
almost an order of magnitude larger than for the large loop, creating
a level of heating that puts it outside the TRACE 171
filter, and thus renders it invisible in \Fig{trace171}. Bright loops
seen in the TRACE 171 filter generally have small winding numbers.

This suggests a scenario where loops are for a short period of time
subjected to excess winding, 
which raises their chromospheric temperatures and evaporates mass
(lowers the local height of the transition zone). The Spitzer
conductivity maintains the coronal part of the loops close to
isothermal, now with increased density and
temperature. If the heating is only moderate, the individual loop may
show up as bright in the TRACE 171 filter for a prolonged period, while if
the heating is large the loop will pass through the TRACE 171 filter
quickly, and will only be visible for a brief period of time.  During this
time these loops should
be characterized by having large velocities along them, because of
un--balanced pressure gradients.

Inspection of the arrangement of field lines along loops reveals an
effect that is a likely explanation for the apparent lack of expansion
with height of the slender loops.  We find that a circular cross section
at the top of loops is mapped to very flat cross sections in the loop
legs, presumably caused by the shearing motions in the photosphere.  This,
or even more complicated arrangement of field lines, is likely to explain
why projected cross sections do not follow the naive $B^{-1/2}$ scaling
expected for cross sections that remain circular.

\ifnum\astroph=1
\FIGIIIem
\fi

\section{Conclusions}

These initial investigations, to be followed by more detailed and
extensive numerical experiments, have established that the DC
(braiding) heating mechanism, originally proposed by \citet{Parker72},
is effective and seems to be sufficient to heat the solar corona.
Because of the very low plasma beta in the low corona above active
regions, the dissipation of even a very small fraction of the
non-potential magnetic energy is sufficient to heat the tenuous
coronal plasma.  The heating has a scale height behavior, consistent
with the observational limits set by \citet{Aschwanden01b} for large
scale coronal loops and by \citet{Aschwanden01a} for a range of loop
sizes.  

The results are not directly comparable to typical active regions on the
Sun, for three related reasons. First, the simulated region is small compared to
typical solar active regions, which makes the loops correspondingly shorter.
Second, the total simulated time is only about 40 minutes, corresponding
to only a few turn-over times of the granular size part of the photospheric
velocity field. This means that motions on larger
scales, with much longer turn-over times, have yet only had little influence
on the magnetic field, so large scale shears are still missing.  Third,
there is no emerging flux at the lower boundary, so part of the Poynting
flux through the lower boundary is missing.  Nevertheless, the results
are tantalizing, and allows the identification of qualitative effects
that are independent of these quantitative shortcomings.

The heating process automatically creates slender loops
consistent with the ones seen in the solar corona by TRACE. These
loops have, at the time when they show up in the emulated TRACE
filter, almost constant temperature and density, and are not in
hydrostatic equilibrium.  The near uniform temperature is caused
by the Spitzer conductivity forcing the loop to keep a small
temperature gradient.  The non-hydrostatic stratification is a
signature of the fact that these loops are dynamic; they are
caused by short duration excess heating, which causes up flows
that increase the density along those magnetic field lines that
are subjected to the excess heating.  This {\rm defines} the
loop, and causes a selection effect; a loop will typically be
observed when its density is near maximum, at which time it is by
definition not in hydrostatic equilibrium.  Even though the loops
modeled here are shorter than the ones where deviations from
hydrostatic equilibrium have been found observationally
\citep{Aschwanden01a}, the same mechanism applies.

\acknowledgments{
BVG acknowledges support through an EC-TMR grant to the European
Solar Magnetometry Network.The work of {\AA}N was supported in
part by the Danish Research Foundation, through its establishment
of the Theoretical Astrophysics Center. Computing time at the
Center for Parallel Computers, was provided by the Swedish
National Allocations Committee.
Both authors gratefully acknowledge the hospitality of LMSAL and ITP/UCSB
(through NSF grant No. PHY99-07949) during this work.}

\ifnum\astroph=1
\end{document}
\fi

\clearpage

\FIGI

\clearpage

\FIGII

\clearpage

\FIGIII


\begin{thebibliography}{}

\bibitem[Aschwanden(2001)]{Aschwanden01c}Aschwanden, M.J. 2001, \apj,
  560, 1035

\bibitem[Aschwanden, Nightingale \&
  Alexander (2000)]{Aschwanden01b}Aschwanden, M.J., Nightingale,
R.W., \& Alexander, D. 2000, {\apj}, 541, 1059

\bibitem[Aschwanden, Schrijver \&
  Alexander (2001)]{Aschwanden01a}Aschwanden, M.J., Schrijver, C.J., \&
  Alexander, D. 2001, \apj, 550, 1036

\bibitem[Aschwanden et al.(2000)]{Aschwandenetal00}Aschwanden, M.J.,
  Tarbell, T.D., Nightingale, R.W., Shcrijver, C.J., Title, A.M.,
  Kankelborg, C.C., Martens, P., \&  Warren, H.P. 2000, \apj, 535, 1047

\bibitem[van Ballegooijen(1986)]{vanBall86}van Ballegooijen,
  A.A. 1986, \apj, 311, 1001

\bibitem[Berger et al. (1998)]{berger98}Berger, T.E., L{\"o}fdahl,
  M.G., Shine, R.A., \& Title, A.M. 1998, \apj, 506, 439

\bibitem[Galsgaard \& Nordlund(1996)]{Galsgaard+Nordlund95xc}
  {Galsgaard, K., \& Nordlund, {\AA}.} 1996
  {Journal of Geophysical Research\/}
  {101}, 13445--13460.

\bibitem[Gomez et al.(2000)]{Gomez+00}Gomez, D.O., Dmitruk, P.A., \&
  Milano, L.J. 2000, \solphys, 195, 299

\bibitem[Hathaway et al.(2000)]{hathaway00}Hathaway, D.H., Beck, J.G.,
  Bogart, R.S., Khatri, G., Petitto, J.M., Han, S., \& Raymond,
  J. 2000, \solphys, 193, 299

\bibitem[Hendrix et al.(1996)]{Hendrix+96}Hendrix, D.L., van Hoven,
  G., Miki{\'c}, Z., \& Schnack, D.D. 1996, \apj, 470, 1192

\bibitem[Heyvaerts \& Priest(1992)]{Heyvaerts+Priest92}Heyvaerts,
  J., \& Priest, E.R. 1992, \apj, 390, 297

\bibitem[Kahn(1976)]{Kahn76}Kahn, F.D., 1976, \aap, 50, 145

\bibitem[Longcope \& Sudan(1994)]{Longcope+Sudan94}Longcope, D.W., \&
  Sudan, R.N. 1994, \apj, 437, 491

\bibitem[Miki{\'c} et al.(1989)]{Mikic+ea89}Miki{\'c}, Z., Schnack,
  D.D., \& van Hoven, G. 1989, \apj, 338, 1148

\bibitem[Nordlund \&
  Galsgaard(1995)]{Nordlund+Galsgaard95mhd}Nordlund, {\AA}., \&
  Galsgaard, K., 1995,
  technical report, Astronomical Observatory, University of Copenhagen

\bibitem[Parker(1972)]{Parker72}
  Parker, E.N. 1972, \apj, 174, 499

\bibitem[Parker(1983)]{Parker83}Parker, E.N. 1983, \apj, 264, 642

\bibitem[Parker(1988)]{Parker88}Parker, E.N. 1988, \apj, 330, 474

\bibitem[Parnell \& Jubb(2000)]{Parnelljubb00}Parnell, C.E., \& Jubb,
  P.E. 2000, \apj, 529, 554

\bibitem[Schrijver et al.(1999)]{Schrijver99}Schrijver, C.J., et al. 1999,
  {\solphys},187, 261

\bibitem[Schrijver \&  Zwaan(2000)]{schrijverzwaan00}
Schrijver, C.J., \& Zwaan, C. 2000, Solar and Stellar Magnetic
Activity, (Cambridge: University Press)

\bibitem[Shine, Simon \& Hurlburt(2000)]{shine00}Shine, R.A., Simon,
  G.W., \& Hurlburt, N.E. 2000, \solphys,193,393

\bibitem[Spitzer(1956)]{spitzer56} Spitzer, L., Jr. 1956,in Physics of fully
        ionized gases,(London: Interscience Publishers Ltd)

\bibitem[Stein \& Nordlund(1998)]{stein98}Stein, R.F., \& Nordlund,
  {\AA}. 1998, \apj, 499,914

\bibitem[Stein \& Nordlund(2000)]{Stein+Nordlund99}Stein, R.F., \&
  Nordlund, {\AA}. 2000, \solphys, 192, 91

\bibitem[Sturrock \& Uchida(1981)]{Sturrock+Uchida81}Sturrock, P.A.,
  \& Uchida, Y. 1981, \apj, 246, 331

\bibitem[Vernazza, Avrett \& Loeser(1981)]{VAL81}Vernazza, J.E., Avrett, E.H.,
\& Loeser, R. 1981, \apjs, 45, 635

\end{thebibliography}
\end{document}